\documentclass[aps,twocolumn,groupedaddress]{revtex4}
\usepackage{epsfig}
\usepackage{graphicx}

\newcommand{\ba}{\begin{eqnarray}}
\newcommand{\ea}{\end{eqnarray}}
\newcommand{\beq}{\begin{equation}}
\newcommand{\eeq}{\end{equation}}
\newcommand{\g}{\gamma}

\def\e{\epsilon}

\begin{document}
\bibliographystyle{apsrev}

\title{Diffuse Galactic Gamma Rays from Shock-Accelerated Cosmic Rays}

\author{Charles D. Dermer}
\affiliation{Space Science Division, 
Code 7653, Naval Research Laboratory, Washington, DC 20375-5352; charles.dermer@nrl.navy.mil}

\date{\today}

\begin{abstract}
A shock-accelerated particle flux $\propto p^{-s}$, where $p$ is the particle momentum,
follows from simple theoretical considerations of cosmic-ray acceleration at 
nonrelativistic shocks followed by rigidity-dependent escape into the 
Galactic halo. A flux of shock-accelerated cosmic-ray protons with $s \approx 2.8$
provides an adequate fit to the {\it Fermi}-LAT $\gamma$-ray emission spectra
of high-latitude and molecular cloud gas when uncertainties in nuclear production models
are considered. 
A  break in the spectrum of cosmic-ray protons claimed
by Neronov, Semikoz, \& Taylor ({\it PRL}, {\bf 108}, 051105, 2012) when fitting
the $\gamma$-ray spectra of high-latitude molecular clouds is a 
consequence of using a cosmic-ray proton flux described by 
a power law in kinetic energy. 
\end{abstract}


\maketitle

{\it Introduction.}---One hundred years after the discovery of cosmic rays by Victor Hess in 1912
\citep{2010EPJH...35..309C}, the sources of the cosmic radiation 
are still not conclusively established.
Theoretical arguments 
developed to explain extensive observations
of cosmic rays and Galactic radiations
favor the hypothesis that cosmic-ray acceleration 
takes place at supernova remnant (SNR) shocks \citep{gs64}. 
A crucial prediction that follows from this hypothesis is that
 cosmic-ray sources will glow in the light of
$\gamma$ rays made by the decay of 
neutral pions created as secondaries in collisions
between cosmic-ray protons and ions with ambient matter. 
The $p+p\rightarrow\pi^0\rightarrow 2\gamma$ spectrum formed by 
isotropic cosmic rays interacting with particles at
rest is hard and symmetric about $\epsilon_\gamma = m_{\pi^0}/2 = 67.5$ MeV
in a log-log representation of 
photon number spectrum vs.\ energy \citep{ste71}.

The {\it AGILE} and {\it Fermi} $\gamma$-ray telescopes
have recently provided preliminary evidence for a $\pi^0\rightarrow 2\gamma$  
feature in the spectra of the W44, W51C, and IC 443 SNRs \citep{2011ApJ...742L..30G}.
Whether this solves the cosmic-ray origin problem depends 
on disentangling leptonic and hadronic emission signatures, 
including multi-zone effects \citep{2012ApJ...749L..26A}, and finding out if the 
nonthermal particles found in middle-aged SNRs as inferred from their spectral maps
have the properties
expected from the sources of the cosmic rays.

Supporting evidence that cosmic rays are accelerated at shocks 
comes from the diffuse Galactic $\gamma$-ray emission. 
High-latitude Galactic gas separate from $\gamma$-ray point sources, dust and 
molecular gas provides a ``clean" target for $\gamma$-ray production from cosmic-ray interactions. Because the cosmic-ray electron
and positron fluxes are $\gtrsim 20$ times smaller than the cosmic-ray 
proton flux at GeV energies \citep{ms98}, the $\pi^0$-decay $\gamma$-ray flux
strongly dominates the electron bremsstrahlung and Compton fluxes. The diffuse 
Galactic $\gamma$-ray spectrum can be deconvolved to give the cosmic-ray 
proton spectrum, given accurate nuclear $\gamma$-ray production physics.
Data from the {\it Fermi} Large Area Telescope (LAT) offer an opportunity to 
derive the interstellar cosmic-ray spectra unaffected by Solar modulation \citep{2009ApJ...703.1249A,2012PhRvL.108e1105N}. 

This problem is revisited in order to address a recent claim \citep{2012PhRvL.108e1105N} 
that fits to the diffuse Galactic $\gamma$-ray emission of Gould belt clouds
imply a break at $T_{k,br}=9^{+3}_{-5}$ GeV  in the cosmic-ray proton spectrum that represents
a new energy scale. Note that a break in the kinetic-energy representation of the proton spectrum
has been reported before \citep{2000ApJ...537..763S}.
In this Letter we show that when uncertainties in nuclear
production models are taken into account, a power-law momentum spectrum favored by cosmic-ray
acceleration
theory provides an acceptable fit to {\it Fermi}-LAT $\gamma$-ray data 
of diffuse Galactic gas, and produces a break in a kinetic energy representation
at a few GeV from elementary kinematics. Comparison of theoretical $\gamma$-ray spectra
with data supports a nonrelativistic shock origin of the cosmic radiation,
consistent with the SNR hypothesis. 

{\it Production spectrum of $\gamma$-rays from cosmic-ray collisions.}---The $\gamma$-ray production 
spectrum divided by the hydrogen density $n_H$ is given, in units of
(s-GeV)$^{-1}$, by
\beq
{F_{pH}(\e_\g)\over n_{H}}= 4\pi k \int_{T_{p,{\rm min}}(\e_\g)}^\infty dT_p\; 
j_p(T_p)\;{d\sigma_{pp\rightarrow\pi^0\rightarrow 2\gamma}(T_p,\e_\g )\over d\e_\g }\;.
\eeq
Here and below, $j(T_p)$ is the cosmic-ray proton intensity in units of cosmic-ray 
protons (cm$^{2}$-s-sr-GeV)$^{-1}$, 
and $k$, the nuclear enhancement factor,
 corrects for the composition of nuclei heavier than hydrogen in the cosmic rays and 
target gas \citep{2009APh....31..341M}.
The term $d\sigma_{pp\rightarrow \pi^0\rightarrow 2\gamma}(T_p,\e_\g )/ d\e_\g$ is the differential cross section 
for the production of a photon with energy $\e_\g$ by a proton with kinetic energy $T_p$, in GeV, and 
momentum $p$ in GeV/c.

The much studied and favored model for cosmic-ray acceleration is the first-order Fermi mechanism, which 
was proposed in the late 1970s \citep{shock}. Test
particles gain energies by diffusing back and forth across a shock front while convecting downstream. 
The downstream steady-state distribution function is given by $f(p)\propto p^{-3r/(r-1)}$,
where $p$ is the momentum and $r$ is the compression ratio. Consequently, the
particle momentum spectrum   $\propto p^2 f(p)\propto p^{-A}$, where $A = (2+r)/(r-1)$ is the
well-known test-particle spectral index that approaches  $2$ (equal energy per decade) in the 
limit of a strong nonrelativistic shock with $r \rightarrow 4$ \cite{1987PhR...154....1B}. After injection into the interstellar 
medium with $A\cong 2.1$ -- 2.2, characteristic of SNR shocks, cosmic-ray protons and ions are transported from the galactic disk into the halo by rigidity-dependent
escape \citep{1979ApJ...229..747J}, which softens their spectrum by $\delta \approx 0.5$ -- 0.6 units, leaving a steady-state 
cosmic-ray spectrum $dN/dp \propto p^{-s}$, where $s = A +\delta\cong 2.7$ -- 2.8. 
Thus we consider a cosmic-ray flux
\beq
j_{sh}(T_p) \propto \beta(T_p)  ({dN\over dT_p}) \propto \beta(T_p)  |{ dp\over dT_p} |\, ({dN\over dp}) \propto p(T_p)^{-s}\;.
\label{jTp}
\eeq

{\it Gamma-ray production from cosmic-ray/matter interactions.}---Secondary nuclear production  
 in proton-proton collisions is described by
isobar formation at energies near threshold $T_p = 0.28$ GeV \citep{{ste71}}, and scaling models 
 at high-energies $T_p \gg m_p$.  Uncertainties remain in 
the production spectra at $T_p\approx$ few GeV, where most of the  
secondary $\gamma$-ray energy is made in cosmic-ray collisions. 

\begin{figure}
{\includegraphics[width=\columnwidth]{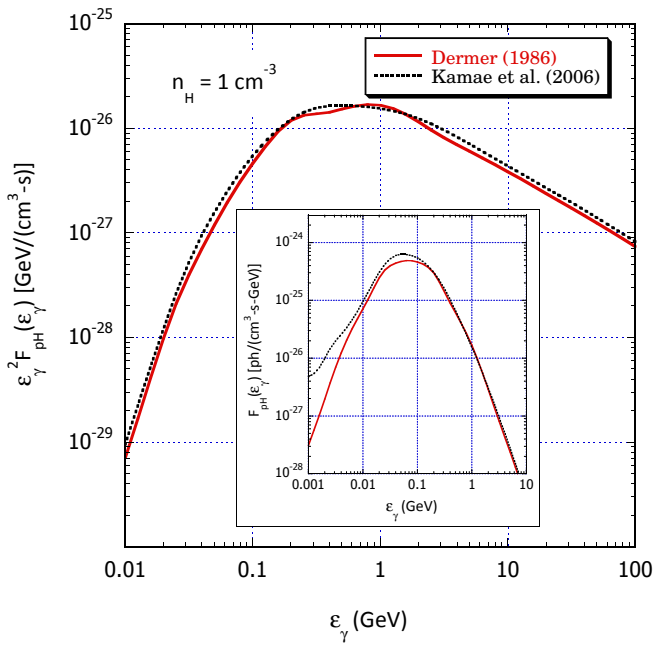}}
\caption[]{
Production spectra of secondary $\gamma$-rays 
made in p-p collisions from the models of Dermer \citep{der86} and Kamae et al.\  \citep{kamae2006},
using the empirical demodulated cosmic-ray proton spectrum given by eq.\ (\ref{jTpder}). 
 }
 \label{Fig1}
\end{figure}


Fig.\ 1 compares two models for $\gamma$-ray production from p-p collisions
using an empirical fit to the demodulated cosmic-ray flux meausred 
\citep{1983ARNPS..33..323S} in interstellar space given by
\beq
j_{dem}(T_p) = 2.2 (T_p+m_p)^{-2.75}\;
\label{jTpder}
\eeq
\citep{der86}. The model of Dermer \citep{der86} describes low-energy resonance production in 
terms of the $\Delta(1232)$ resonance through Stecker's model \citep{ste71}, and high-energy production by the 
scaling model of Stephens 
and Badhwar \citep{1981Ap&SS..76..213S}, with the two regimes linearly connected between
3 and 7 GeV. The model of Kamae et al.\ \citep{kamae2006} is a parametric
 representation of simulation programs, and includes contributions
from the $\Delta(1232)$ isobar and $N(1600)$ resonance cluster, non-scaling effects, 
 scaling violations, and diffractive processes. This model is represented by 
functional forms that are convenient for astrophysical calculations.

Fig.\ \ref{Fig1} shows that the two models agree within 20\% at $\epsilon_\gamma > 100$ MeV, 
but display larger differences at $\epsilon_\gamma \lesssim 10$ MeV,  where the $\gamma$-ray production
is energetically insignificant.
Both models are in general agreement in the high-energy asymptotic regime,  with the 
 Kamae et al.\ model giving fluxes larger by $\approx 13$\% 
due to the inclusion of diffractive  processes. 
 The spectral peak in a $\nu F_\nu$ representation occurs at 500 MeV -- 1 GeV for
this proton spectrum. The most significant discrepancy in the two models is by $\approx 30$\% at the 
pion production peak near 67.5 MeV \cite{2011arXiv1106.5073C}. Because of the different approaches of the models, 
and little improvement in nuclear data bases from laboratory studies between model development,
this comparison indicates that our knowledge of the $\gamma$-ray production spectrum in p-p
collisions is uncertain, at worst, by 30\% near the pion-production peak, and is better than 
15\% at $\epsilon_\gamma \gtrsim 200$ MeV.

\begin{figure}
{\includegraphics[width=\columnwidth]{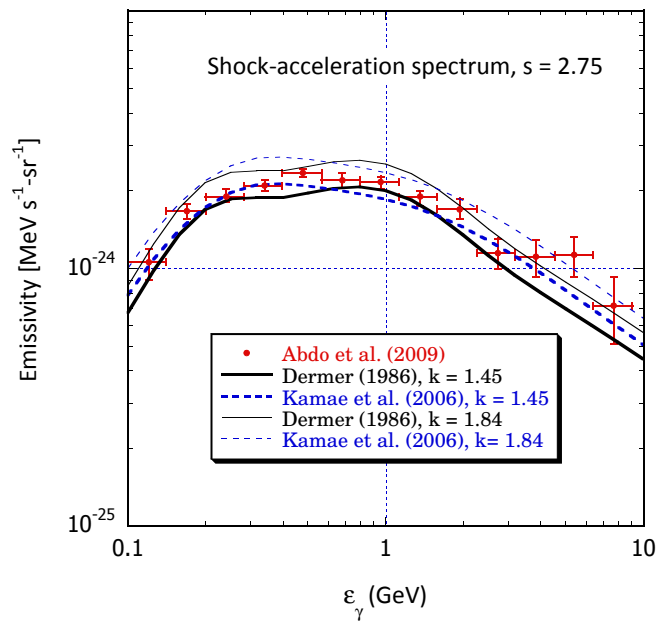}}
\caption[]{
Fits to the {\it Fermi}-LAT spectrum of the differential  $\gamma$-ray
emissivity of local neutral gas \citep{2009ApJ...703.1249A}, employing a shock-acceleration spectrum,
Eq.\ (\ref{jCR}), with $s = 2.75$, for 
the cosmic-ray proton spectrum, and the models of Dermer \citep{der86} and Kamae et al.\  \citep{kamae2006}
for $\gamma$-ray production. 
The nuclear enhancement factor $k$ takes the value of 1.45 and 1.84, as labeled.
 }
 \label{Fig2}
\end{figure}

\begin{figure}
{\includegraphics[width=\columnwidth]{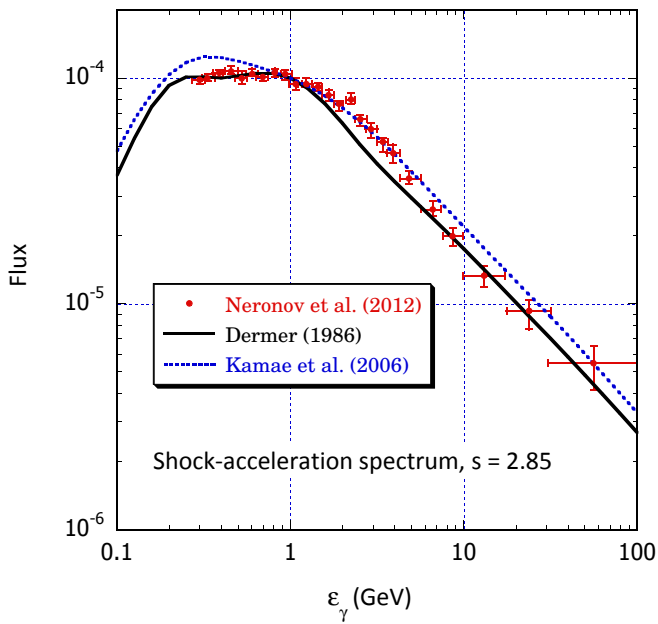}}
\caption[]{
 {\it Fermi}-LAT spectrum of the differential  $\gamma$-ray
emissivity of Gould belt clouds  \cite{2012PhRvL.108e1105N}. It is fit with a shock-acceleration model for 
the cosmic-ray proton spectrum with $s = 2.85$, using the models of Dermer \citep{der86} and Kamae et al.\  \citep{kamae2006}
for $\gamma$-ray production. The models are normalized to the flux at 1 GeV.
 }
 \label{Fig3}
\end{figure}

{\it Fits to {\it Fermi}-LAT $\gamma$-ray data}---The Fermi-{\it LAT} data \citep{2009ApJ...703.1249A} of the diffuse galactic $\gamma$ radiation are fit using a shock spectrum given
by Eq.\ (\ref{jTp}). Guided by the high-energy asymptote of 
Eq.\ (\ref{jTpder}), we let 
\begin{equation}
j_{CR}(T_p) = 2.2 \,p^{-s} \;,
\label{jCR}
\end{equation}
with $s = 2.75$. The data shown in Fig.\ \ref{Fig2} are from regions in the third quadrant, with 
Galactic longitude from 200$^\circ$ to 260$^\circ$, and galactic latitudes $|b|$ ranging from 22$^\circ$ to 60$^\circ$. 
There are no molecular clouds in these regions, the ionized hydrogen
column density is small with respect to the column density of neutral hydrogen, and $\gamma$-ray emission from 
point sources is removed. The linear increase of the 
emissivity as traced by 21 cm line observations allows residual galactic and extragalactic $\gamma$ radiation to 
be subtracted, leaving only the $\gamma$-ray emission resulting collisions of cosmic rays with neutral gas, 
allowing for an absolute normalization to be derived for the emissivity.  As can be seen, the shock-acceleration spectrum
gives an acceptable fit to the data, and restricts the value of $k$ to be $\lesssim 1.8$. Given that there must be some residual 
cosmic-ray electron-bremsstrahlung and Compton radiations at these energies, the restriction on $k$ could be larger.

\begin{figure}
{\includegraphics[width=\columnwidth]{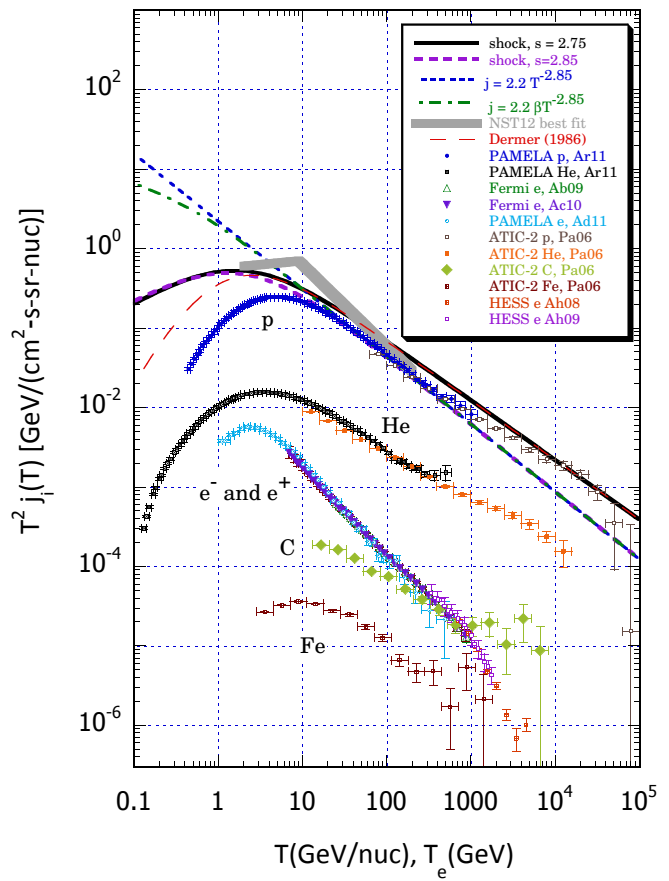}}
\caption[]{
Theoretical cosmic-ray proton fluxes 
compared with recent PAMELA, {\it Fermi}-LAT, ATIC-2, 
and HESS measurements of cosmic-ray 
p, He, C, Fe, and electron and positron fluxes \citep{CRflux}.
The shock acceleration spectra are given by Eq.\ (\ref{jCR}) with $s = 2.75$ and 
$s=2.85$, a power-law kinetic energy flux $j\propto T^{-2.85}$, and a flux 
$j\propto \beta T^{-2.85}$ are shown, in addition to   
the demodulated cosmic-ray proton spectrum, eq.\ (\ref{jTpder}), and
the best fit spectrum of Ref.\ \cite{2012PhRvL.108e1105N}. Solar modulation
accounts for the difference between the 
measured and theoretical cosmic-ray proton fluxes at $T \lesssim 10$ GeV/nuc. 
Papers cited in \cite{CRflux} report systematic errors for different experiments.
 }
 \label{Fig4}
\end{figure}

The spectrum in Fig.\ \ref{Fig3} from the analysis of {\it Fermi}-LAT data given 
in Ref.\ \cite{2012PhRvL.108e1105N} shows the averaged $\gamma$-ray flux from molecular clouds
in the Gould belt.
The derived flux,
which extends to $\approx 100$ GeV,  matches
the spectrum of diffuse Galactic gas emission used by the {\it Fermi} team in 
their analysis of the extragalactic diffuse $\gamma$-ray intensity \citep{{2010PhRvL.104j1101A}},
supporting the assumption that the clouds are porous to cosmic rays.
The  cosmic-ray proton shock acceleration spectrum given by Eq.\ (\ref{jCR}) with $s = 2.85$ is seen to 
give an 
adequate fit to the data in Fig.\ \ref{Fig3}, given the nuclear physics uncertainties in $\gamma$-ray production.

This  cosmic-ray proton flux is compared in Fig.\ \ref{Fig4} 
with a shocked spectrum given by  Eq.\ (\ref{jCR}) with s= 2.85, with two 
power-law kinetic energy spectra (one multiplied by $\beta$), and the spectrum, eq.\ (\ref{jTpder}), 
assumed to represent the demodulated local cosmic-ray proton spectrum.
The best fit of the functional form used by Neronov et al.\ (2012) \citep{2012PhRvL.108e1105N} is also 
plotted, and is in accord with the shock spectrum, Eq.\ (\ref{jCR}), when allowance is made for the large uncertainties in 
derived parameters. Indeed, the allowed spectrum would be further limited by the {\it Fermi}-LAT 
data of Fig.\ \ref{Fig2} unless $k\lesssim 1.2$. The reduced $\chi^2$ to determine goodness of fit should
take into account uncertainties in nuclear production,  and firm conclusions about a break depend on
improved cross sections.

{\it Summary.}---Fig.\ 4 compares theoretical and empirical local interstellar spectrum of cosmic-ray protons with recent measurements of the fluxes of cosmic-ray protons, He, C, Fe and electrons.  The small fluxes of the heavier cosmic rays and electrons imply that $\gamma$-ray emission from these channels make a minor, but non-negligible contribution to the $\gamma$-ray flux.



The favored model for the origin of cosmic rays is nonrelativistic shock acceleration 
by SNRs in the Galaxy. The simplest possible shock-acceleration model with proton
flux $\propto p^{-s}$ gives an adequate fit to {\it Fermi}-LAT data of high Galactic
latitude gas and clouds, and is in accord with expectations of a SNR origin for the 
cosmic rays. The use of a shock-spectrum for the cosmic-ray flux
allows a range of calculations to be made that depend on knowing the low-energy cosmic-ray spectrum  \citep{2012MNRAS.421L.102O}.
The combination of a power-law momentum injection spectrum and
the observationally similar local interstellar spectrum potentially puts constraints on
propagation models that have a large effect on the primary spectra, such
as those involving strong reacceleration \cite{ms98}.

If cosmic rays are indeed accelerated by SNR shocks, then the $\pi^0\rightarrow 2\gamma$ 
feature from shock-accelerated cosmic rays should be observed in the spectra of individual SNRs.
Indications for such a feature are found in a few middle-aged SNRs \citep{2011ApJ...742L..30G}, 
but final confirmation of cosmic-ray sources will 
require detailed spectral calculations involving both shock-accelerated protons and leptons \citep{2008MNRAS.384.1119F}, 
including 
radiative losses and escape, and comparison with improving {\it Fermi}-LAT data resulting from 
 increasing exposure and development of better analysis tools.

\vskip 0.1in
\noindent
I thank A.\ Atoyan, J.\ D.\ Finke, J.\ Hewitt, R.\ J.\ Murphy, and S.\ Razzaque for discussions, 
and Professor T.\ Kamae for supplying his secondary nuclear production 
code. I would also like to thank A.\ Strong for many interesting and useful
comments, and I.\ Moskalenko and T.\ Porter for pertinent remarks. 
This work is supported by the Office of Naval Research and NASA through the
{\it Fermi} Guest Investigator program.


Note added in proof:
A recent paper by Blasi, Amato, and Serpico [Phys. Rev. Lett. 109, 061101 (2012)] 
explains a supposed low-energy break in a power-law injection momentum spectrum as a 
result of advective effects on cosmic-ray propagation (compare Ref. [13]). 
M. Kachelriess and S.  Ostapchenko (arXiv:1206.4705) claim to find deviations 
from a cosmic-ray power-law momentum  spectrum at low energy,
but their treatment is susceptible to some of the same criticisms as 
presented here.

\end{document}